# Short-Term Forecasting of Thermostatic and Residential Loads Using Long Short-Term Memory Recurrent Neural Networks

Bang Nguyen
*Wind Hybrid Energy System*
*National Renewable Energy Laboratory*
Boulder, CO, USA
bang.nguyen@nrel.gov

Mayank Panwar
*Power System Engineering*
*National Renewable Energy Laboratory*
Boulder, CO, USA
mayank.panwar@nrel.gov

Rob Hovsapian
*ARIES Research Advisor*
*National Renewable Energy Laboratory*
Boulder, CO, USA
rob.hovsapian@nrel.gov

Yashodhan Agalgaonkar
*Power System Engineering*
*National Renewable Energy Laboratory*
Boulder, CO, USA
yashodhan.agalgaonkar@nrel.gov

***Abstract*—Internet of Things (IoT) devices in smart grids enable intelligent energy management for grid managers and personalized energy services for consumers. Investigating a smart grid with IoT devices requires a simulation framework with IoT devices modeling. However, there lack comprehensive study on the modeling of IoT devices in smart grids. This paper investigates the IoT device modeling of a thermostatic load and implements the recurrent neural networks model for short-term load forecasting in this IoT-based thermostatic load. The recurrent neural network structure is leveraged to build a load forecasting model on temporal correlation. The temporal recurrent neural network layers including long short-term memory cells are employed to learn the data from both the simulation platform and New South Wales residential datasets. The simulation results are provided for demonstration.**

## I. Introduction

The electric grid modernization requires more cutting-edge technologies in information than just traditional generation, transmission, and distribution infrastructures. For better monitoring, control, and management of the grid, communication, and control systems play crucial roles [1]. Internet of Things (IoT) technology, which is an emerging technology, enables the integration of information and communication infrastructure to monitoring and controlling devices [2]. IoT enhances the available and reliable data communication for intelligently controlling and managing the grid. IoT devices can also benefit the end-user by providing personalized energy services [3]. Generally, IoT can be integrated into smart meters, electric vehicles, solar photovoltaic panels, and battery energy storage systems, to name a few. Hence, these devices can be involved in supporting the grid.

Analyzing and designing the modern electric grid requires the real-time emulation of power systems with IoT devices [4]. Therefore, modeling IoT devices is significant to evaluate their effects on smart grids. The real-time emulation of a modern grid with IoT devices includes the bi-directional data transfer between digital real-time simulation and hardware under test for power hardware-in-loop (PHIL) and/or controller hardware-in-loop (CHIL). Further, the intelligent control and management strategies for the modern electric grid can be investigated [5]. Modeling an IoT device can be complex since it includes various sensors, actuators, and other physical components downstream [6]. Also, an IoT device has its internal control loops. There are two main approaches to modeling methods. The model-based design focuses on the mathematical equations to describe an IoT device, whereas the model-free one uses statistical learners such as neural networks. Recently, deep learning methods have been successfully applied in the modeling of such a complex system as an IoT device [7].

This paper concerns the modeling of IoT devices, specifically, a residential thermostatic load, under environmental changes. The thermostatically controlled loads contribute 36% of the residential consumption, therefore can involve in demand response [8] for frequency support. This paper investigates the simplified model of the thermostatic load under dynamic environments. While the model still includes the distinct behaviors of such thermostatic loads, it is sufficiently simple to be implemented in an edge device. This paper does not cover the grid-support control of the thermostatic load, but its dynamic load profile under environmental changes. Thereafter, the load forecasting model is built for optimal energy management over IoT-based smart-grid in future works.

The paper implements recurrent neural networks (RNN) with long short-term memory (LSTM) cells for short-term load forecasting. Short-term load forecasting (STLF) focuses on the forecasting of loads from several minutes up to one week into the future [9][10]. Popular methods for load forecasting include autoregressive moving average (ARMA) [11], autoregressive integrated moving average (ARIMA) [12], support vector regression (SVR) [13]. There are several machine learning techniques are introduced for load forecasting [14]. Feed-forward neural network (FFNN) is introduced in [15]. The RNN with its variants such as LSTM and gated recurrent unit (GRU) have been applied for load forecasting in [16], [17]. Short-term load forecasting via graph neural networks has been proposed in [18]. The RNN with LSTM is implemented in this paper for its simple structure and effectiveness.

The rest of this paper is organized as follows. Section II discusses the modeling of IoT-based thermostatic load. Section III describes the RNN with LSTM structure. The load forecasting results are shown in Section IV with data from simulations of the thermostatic load and from New South Wales buildings [14]. Section V concludes the paper.

## II. IoT-Based Thermostatic Load Modeling

This paper focuses on the thermostatic load operated by a relay-based controller in a residential house. This smart home can transact with the energy provider for information exchange and execution of load management commands via IoT technology. Due to the page limit, only this thermostatic load is considered here, other residential loads would be covered in future works.

The block diagram of the IoT-based thermostatic load with the relay-based controller is shown in Fig. 1. The room model describes the thermodynamical behavior of a house in a linear-invariant state-space representation [19] as follows.

$$\begin{aligned} \dot{x} &= Ax + Bu + Ed, \\ T &= Cx, \end{aligned} \tag{1}$$

where the state vector $x$ includes the floor temperature, the internal façade temperature, the external façade temperature, and the indoor temperature, the control input $u$ is the heat flow from the heater, the disturbance vector includes the external temperature, heat from other sources such as occupants or appliances, and the solar radiation, the output $T$ is the indoor temperature. The state-transition matrices employed here are the same as [19] as follows.

$$A = 10^{-3}\begin{bmatrix} -0.020 & 0 & 0 & 0.020 \\ 0 & -0.020 & 0.001 & 0.020 \\ 0 & 0.001 & -0.056 & 0 \\ 1.234 & 2.987 & 0 & -4.548 \end{bmatrix}, \tag{2}$$

$$B = 10^{-3}\begin{bmatrix} 0 \\ 0 \\ 0 \\ 0.003 \end{bmatrix}, D = 10^{-3}\begin{bmatrix} 0 & 0 & 0 \\ 0 & 0 & 0 \\ 0.055 & 0 & 0 \\ 0.327 & 0.003 & 0.001 \end{bmatrix}. \tag{3}$$

The output matrix is $C = [0 \quad 0 \quad 0 \quad 1]$. The initial state is set to $x_0 = [21 \quad 21 \quad 21 \quad 21]^T$. The heater model is the change-rate limiter that converts the ON-OFF signals of 4000 W from the relay-based controller to a ramp of 100 W per second. The relay-based controller provides the ON signal of 4000 W if the indoor temperature $T$ is larger than the user-defined temperature setpoint $s$ plus the tolerance $\gamma$; the OFF signal of 0 W is outputted if $T$ is smaller than $(s-\gamma)$. The control command keep unchanged or selected optimally when $(s-\gamma) \leq T \leq (s+\gamma)$. $s \pm \gamma$ is the comfortable zone of the user-defined temperature. The behaviors under the disturbances of $[-6 \quad 500 \quad 500]$, respectively, of the thermostatic load are shown in Fig. 2.

The IoT interface updates the power consumptions of the residential house to the energy provider and receives the energy commands to economically optimize the power consumption. In this paper, we investigate the load profile of the thermostatic with a non-optimal relay-based controller without energy commands from the energy provider. The load profile is dictated by the distance parameters. The load profile is investigated over a seven-day window in the winter with the disturbance change shown in Fig. 3.

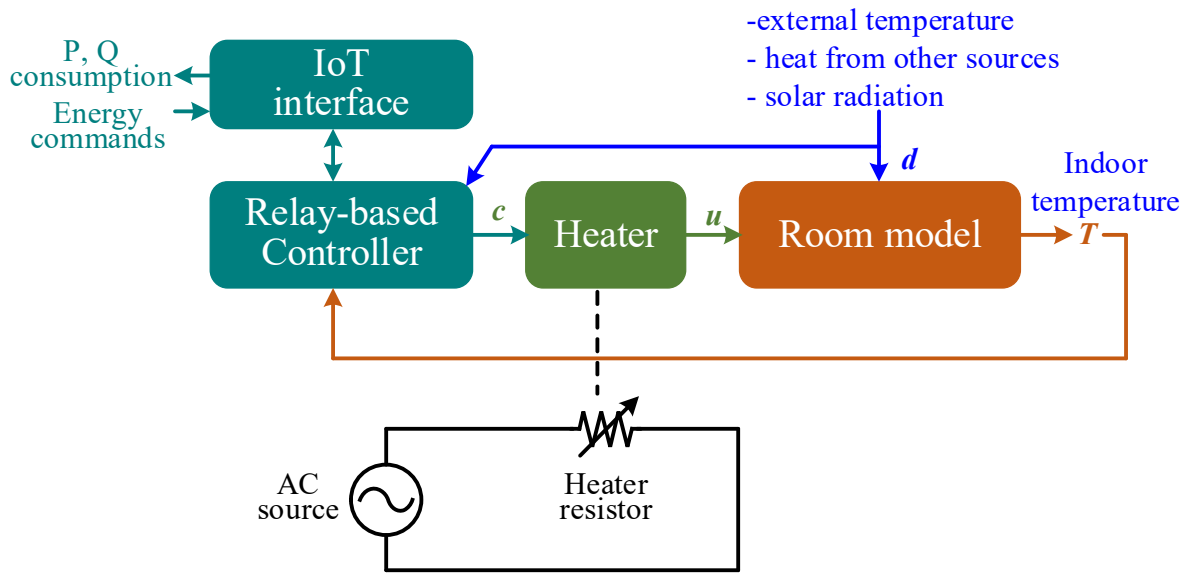

Fig. 1. Block diagram of IoT-based thermostatic load model.

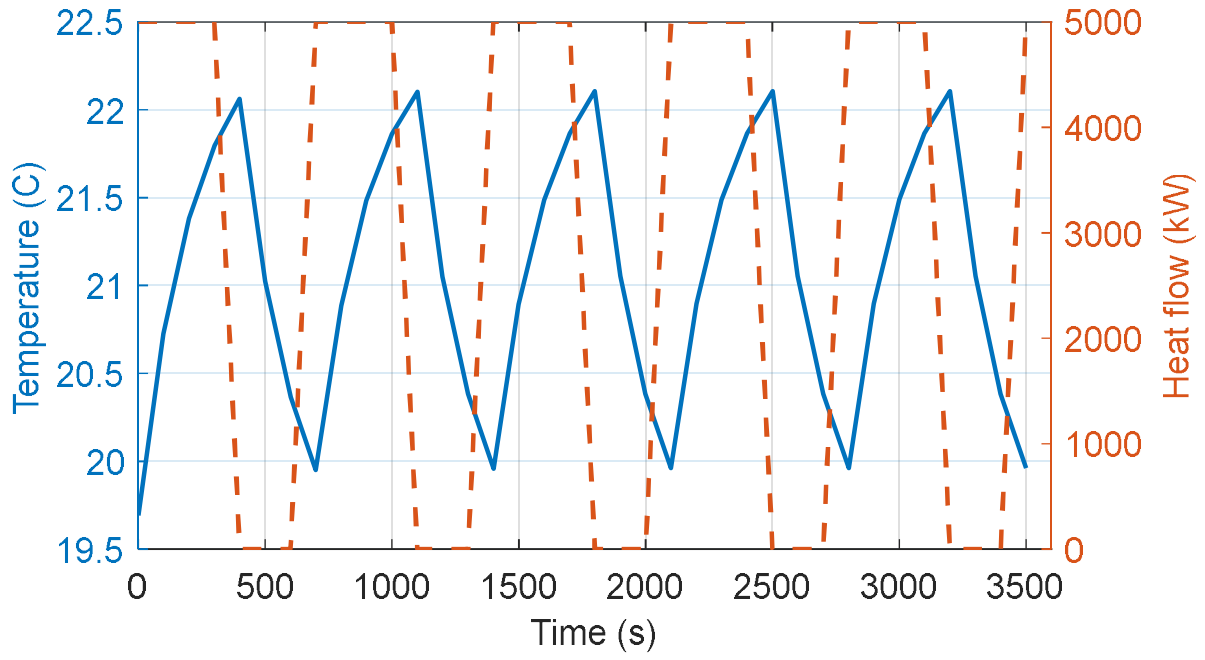

Fig. 2. Indoor temperature response under the switching heat flow.

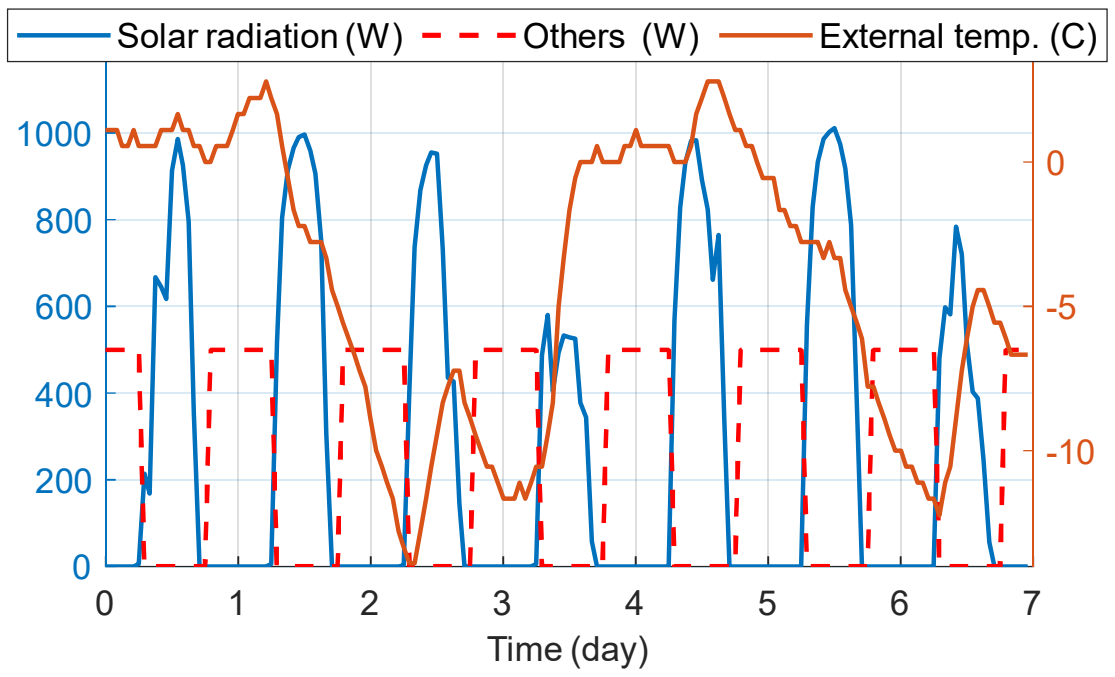

Fig. 3. External disturbance profiles for 7-days.

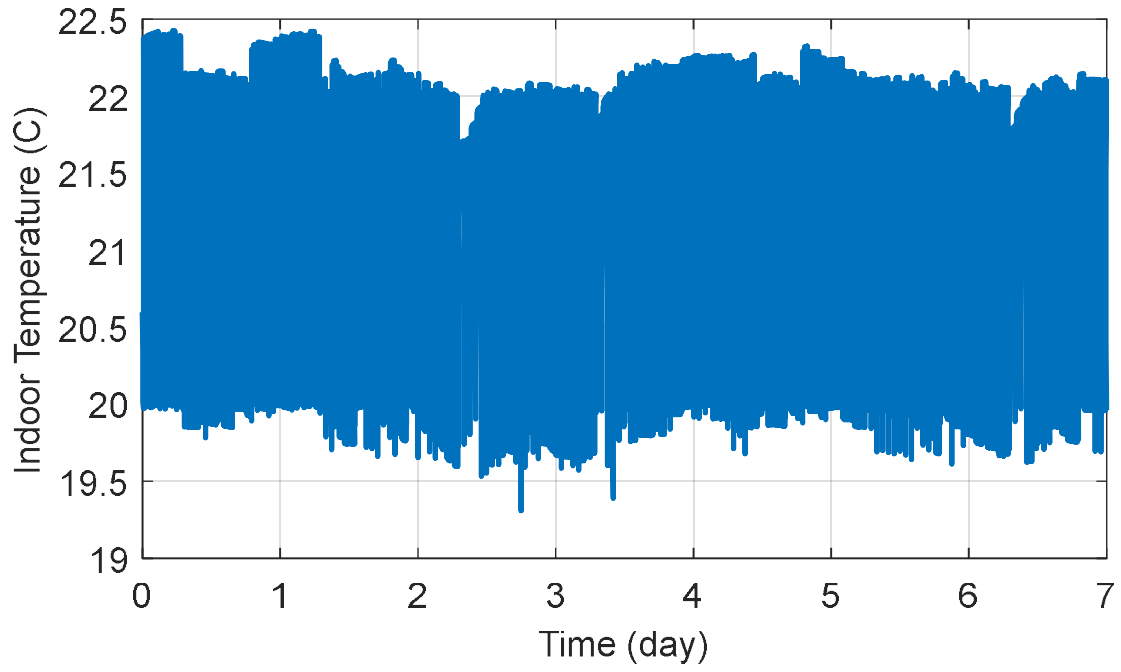

Fig. 4. Indoor temperature response with the close-loop control of the relay-based ON-OFF controller.

Under the close-loop control of the relay-based controller, the indoor temperature is regulated between 21 ºC degrees with a tolerance of 1 ºC as shown in Fig. 4. The heat power consumption over 7 days is shown in Fig. 5 in both switching waveforms and average values. This load profile of the thermostatic load is employed for the load forecasting in Section IV.

## III. Recurrent Neural Networks with LSTM Cell

The RNN with LSTM [20] structure is shown in Fig. 6. One LSTM cell computes for each time step the hidden state $\boldsymbol{h_t}$ and the cell state $\boldsymbol{c_t}$ from the input $\boldsymbol{x_t}$, the previous hidden state $\boldsymbol{h_{t-1}}$, and the previous cell state $\boldsymbol{c_{t-1}}$. In each LSTM, there are intermediate states of the forget gate $\boldsymbol{f_t}$, the cell candidate $\boldsymbol{g_t}$,, the input gate $\boldsymbol{i_t}$, and the output gate $\boldsymbol{o_t}$. The relationship between these state variables is expressed as follows.

$$f_t = \sigma_g\left(W_f x_t + R_f h_{t-1} + b_f\right), \quad (4)$$

$$g_t = \sigma_c\left(W_g x_t + R_g h_{t-1} + b_g\right), \quad (5)$$

$$i_t = \sigma_g(W_i x_t + R_i h_{t-1} + b_i), \quad (6)$$

$$o_t = \sigma_g(W_o x_t + R_o h_{t-1} + b_o). \quad (7)$$

The matrices $W_f$, $W_g$, $W_i$, $W_o$, $R_i$, $R_g$, $R_o$, $R_i$ and the biased vectors $b_f$, $b_g$, $b_i$, $b_o$ are the trainable weights. The gate activation functions $\sigma_g(\cdot)$ is sigmoid, and $\sigma_c(\cdot)$ is tanh. The cell state and hidden state are computed as

$$c_t = f_t \circ c_{t-1} + f_t \circ c_{t-1}, \quad (8)$$

$$h_t = f_t \circ \sigma_c(c_{t-1}), \quad (9)$$

where ∘ denotes the element-wise product.

The RNN LSTM regression network has the input feature of 1 since the input time-series data has 1 dimension.

The LSTM layer is designed to have 200 hidden units. The regression output also is a scalar variable. The implemented RNN LSTM structure is shown in Fig. 7.

## IV. Load Forecasting using LSTM-RNN

The load forecasting is performed on two datasets. The first one is the average heat power consumption of the simulated thermostatic load shown in Fig. 5. The second one is the load consumption on 400 buildings in New South Wales in 2013 [14].

### *A. Heat Power Consumption Forecasting.*

The time-series data of 7 days operations with dynamic environments are employed for this heating power consumption forecasting. The LSTM-RNN model is trained with the data of the first 6 days and provided the prediction on the last day as shown in Fig. 8. The forecasting performance is expressed in Fig. 9. The root-mean-square error (RMSE) is 1293.3424. The 7-day load profile of the thermostatic load is highly fluctuant, so

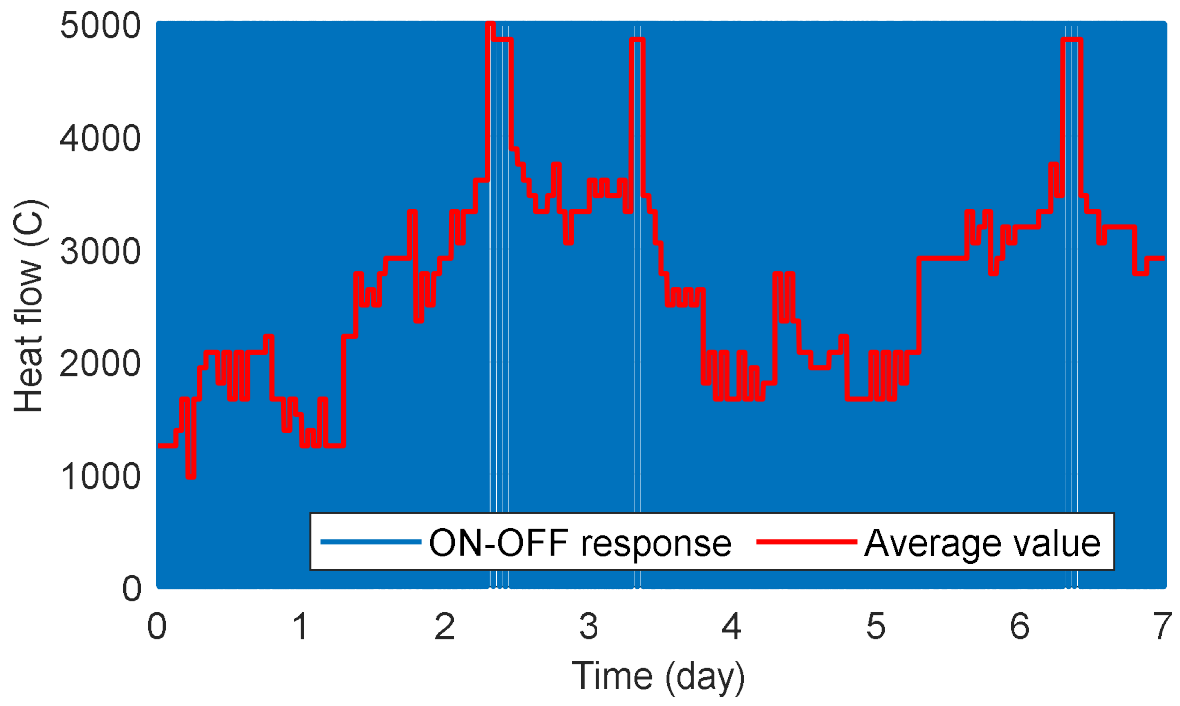

Fig. 5. The heat power consumption in switching waveform and average.

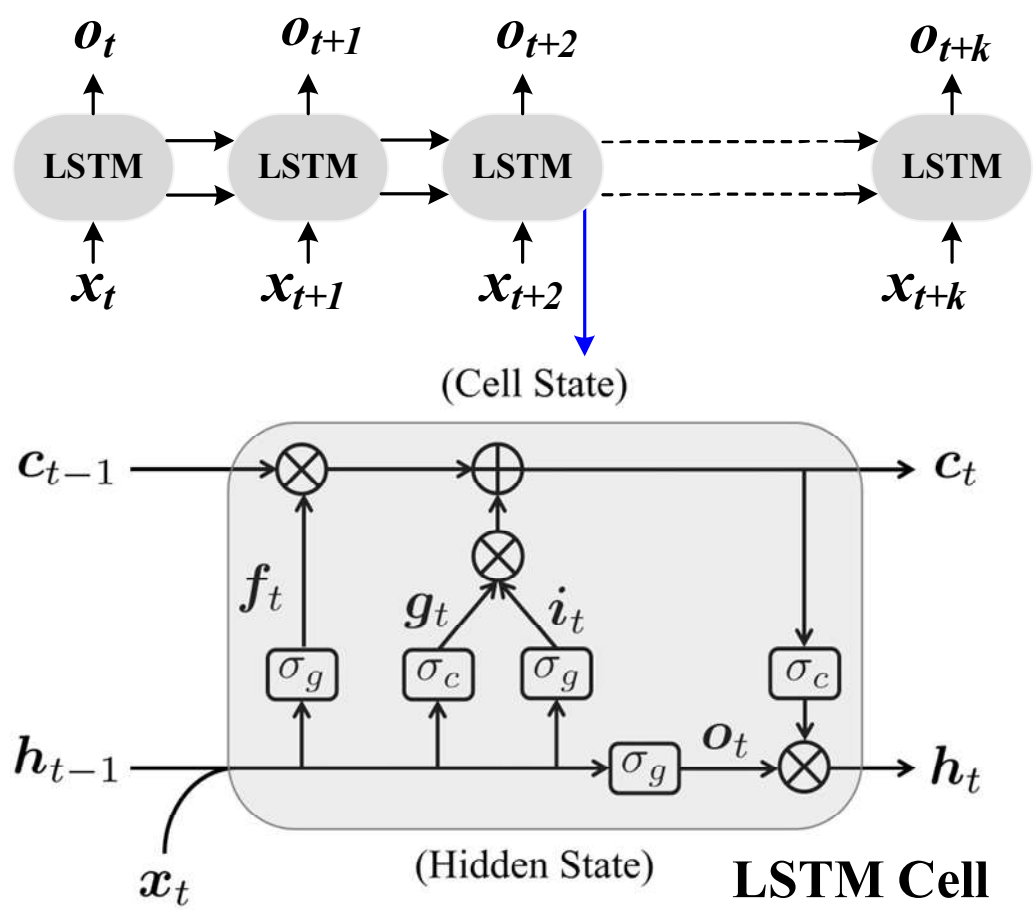

Fig. 6. The general structure of an RNN with LSTM cell.

```
4×1 Layer array with layers:

     1   ''   Sequence Input       Sequence input with 1 dimensions
     2   ''   LSTM                 LSTM with 200 hidden units
     3   ''   Fully Connected      1 fully connected layer
     4   ''   Regression Output    mean-squared-error
```

Fig. 7. The implemented LSTM-RNN structure for load forecasting.

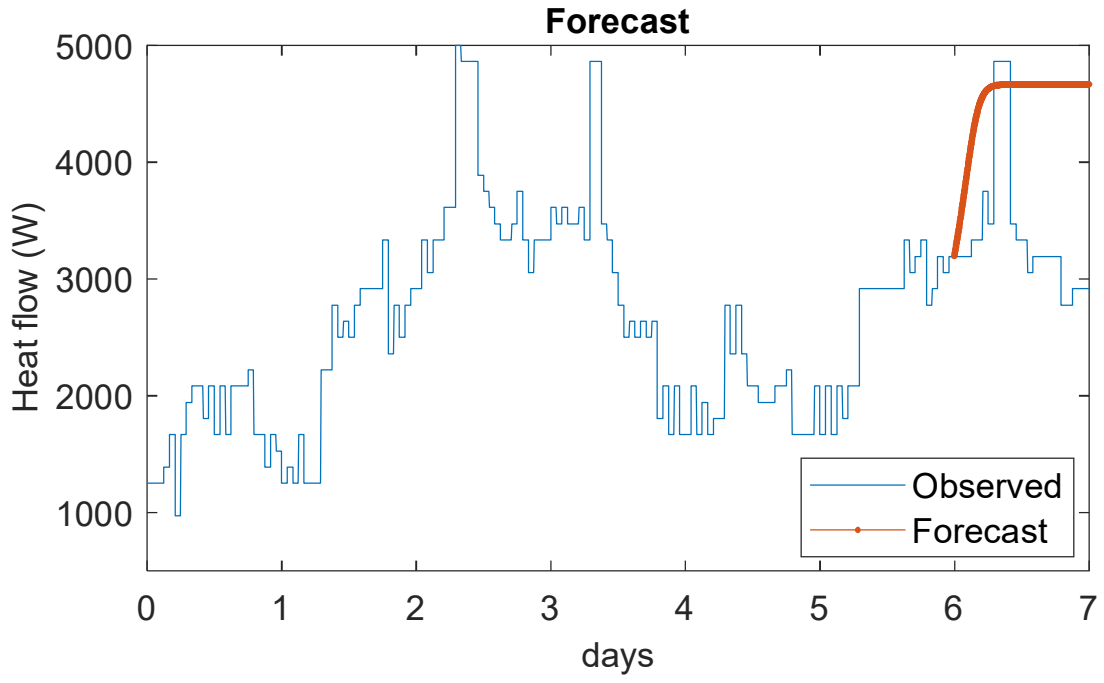

Fig. 8. The load-forecasting of day 7 with time-series data of days 1-6

the predicting values are far from the actual values. As can be seen, the thermostatic load profile is influenced significantly by the environments, so including the prediction of environment parameters can enhance the forecasting performance.

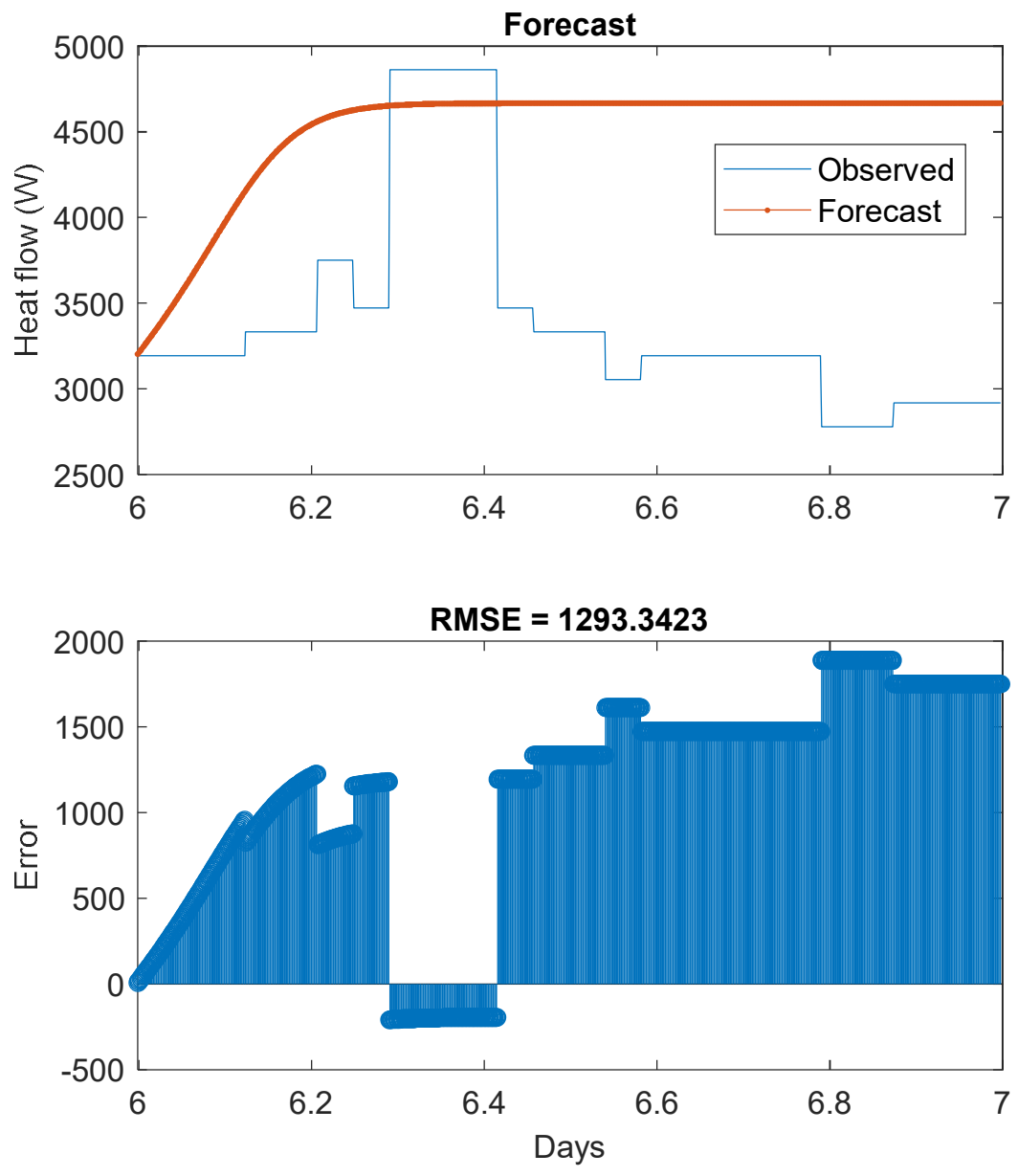

Fig. 9. The 1-day forecasting performance on day 7.

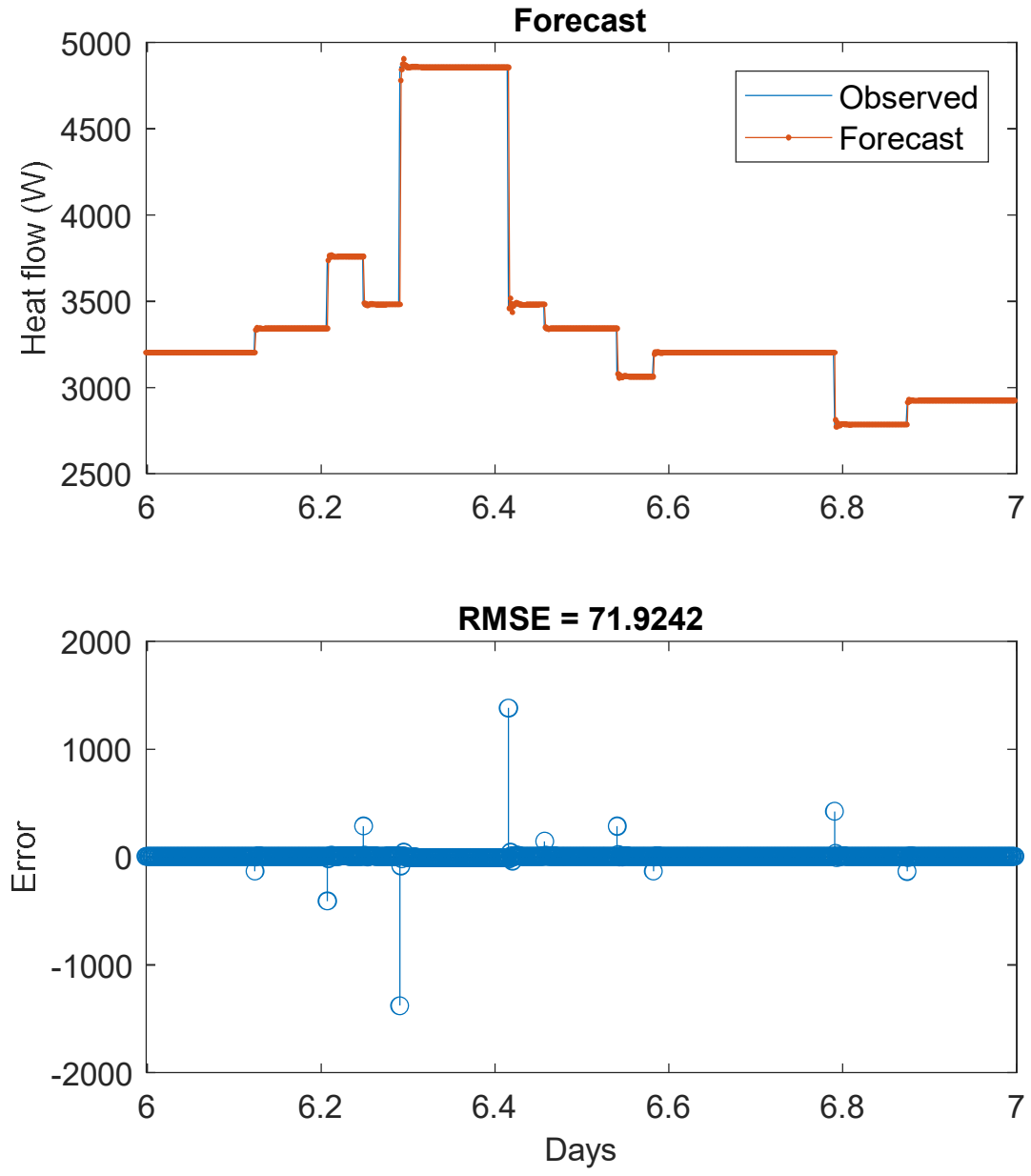

Fig. 10. The one time-step (100s) forecasting performance on day 7.

When having access to the observed data before predictions, we can provide more accurate predictions by reducing the prediction period to only a one-time step. The time-step here is 100 s. The forecasting performance, in this case, is shown in Fig. 10. The LSTM-RNN model can provide relatively accurate prediction values with the RMSE is only 71.9242.

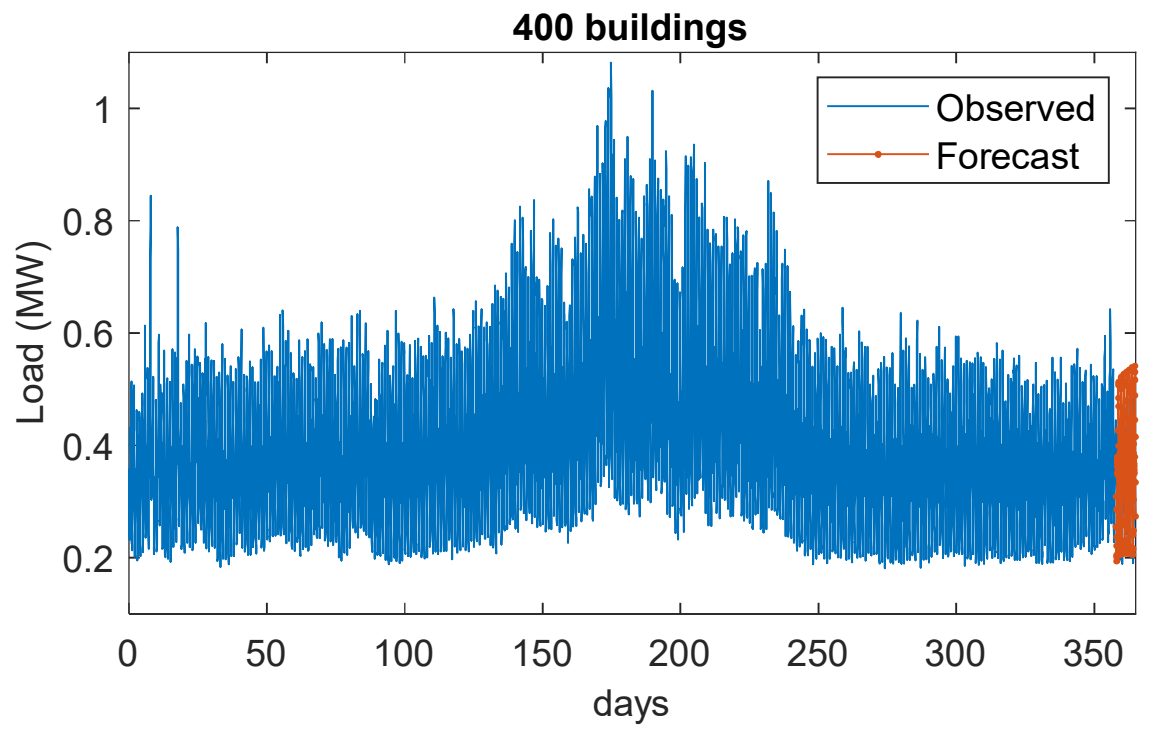

Fig. 11. 400 building load profile and the last 7-day prediction values.

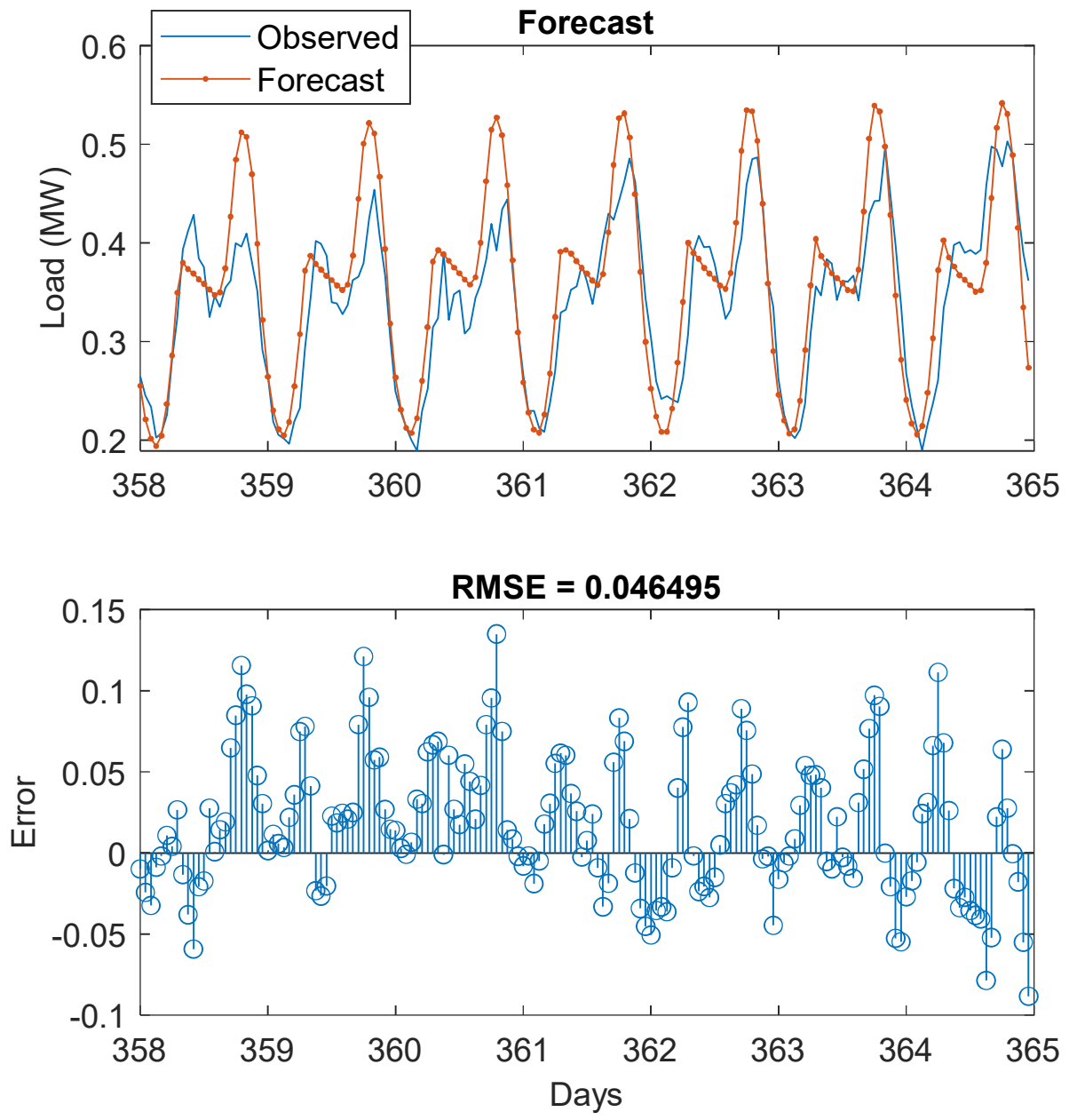

Fig. 12. The 7-day forecasting performance on the last week.

### *A. 400 Buildings Load Demand Forecasting.*

The 365-day time-series data of 400 building load profiles are employed in this case. The LSTM-RNN model is trained with the data of the first 358 days and provides the prediction on the last 7 days as shown in Fig. 11. The forecasting performance is shown in Fig. 12, where the forecasting values are near to the observed ones. The RMSE is only 0.046495.

When having access to the observed data before predictions, we can predict more accurately by reducing the prediction period to only a one-time step. The time-step here is 1 hour. The forecasting performance, in this case, is shown in Fig. 13. The LSTM-RNN model can provide relatively accurate prediction values with the RMSE is only 0.022264.

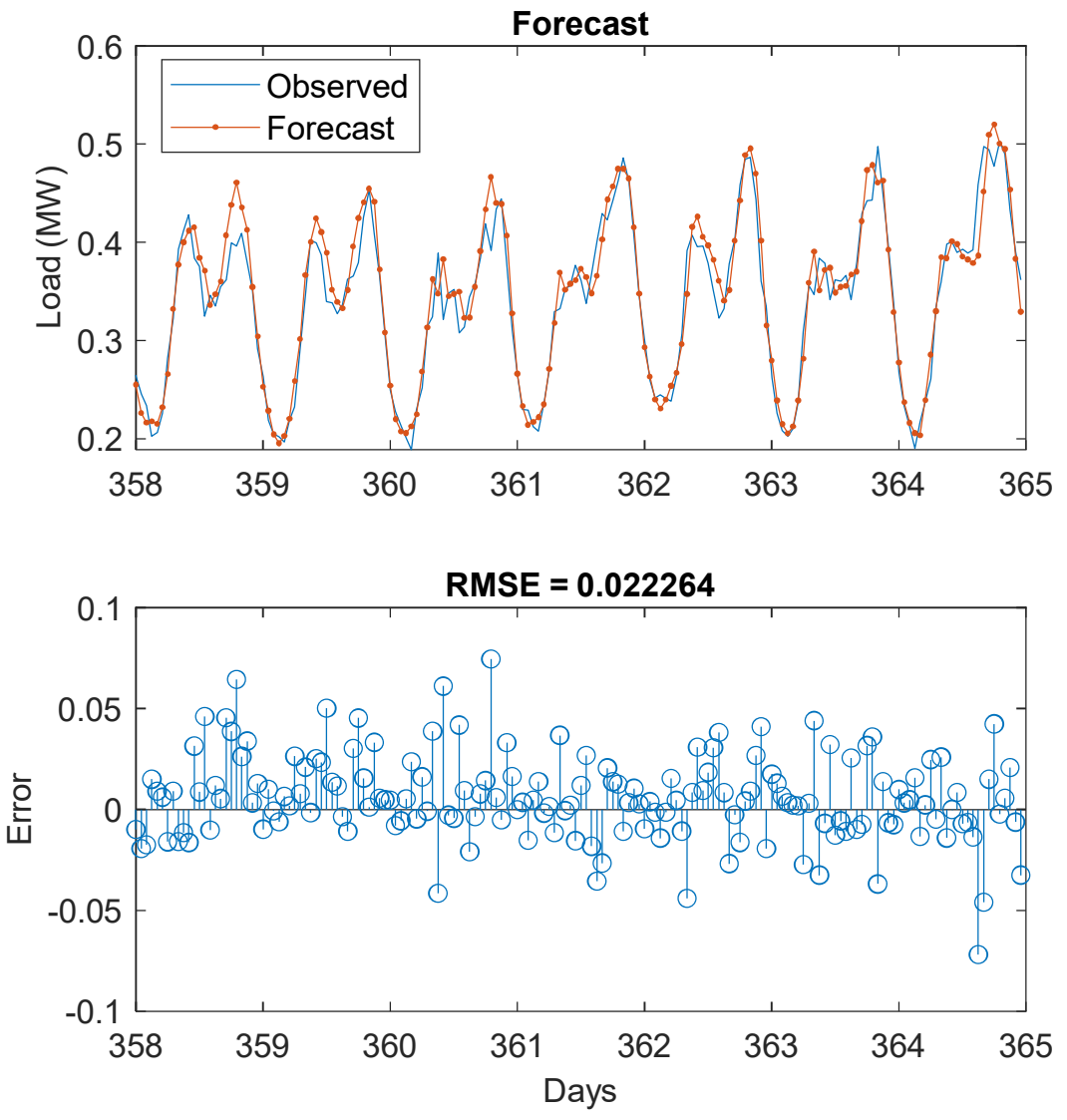

Fig. 13. The one time-step (1 hour) forecasting performance on the last week.

## V. Conclusion

In this paper, we investigate the modeling of a residential thermostatic load under dynamic environments for IoT-based energy management. Further, the short-term load forecasting framework based on LSTM-RNN is implemented for time-series data from the investigated thermostatic load profiles over 7 days. Although the 1-day prediction performance is not high, the one time-step (100 s) prediction is relatively accurate. The LSTM-RNN model is also trained and tested on the 400 buildings load profile. Both 7-day predictions and 1 time-step (1 hour) predictions are quite accurate due to the periodic feature of the load profile. Future work will include the environment parameters on the forecasting model of residential load profiles. More IoT-based residential load is also interesting for modeling and forecasting tasks.

## Acknowledgment

This work was authored in part by the National Renewable Energy Laboratory, operated by Alliance for Sustainable Energy, LLC, for the U.S. Department of Energy (DOE) under Contract No. DE-AC36-08GO28308. The views expressed in the article do not necessarily represent the views of the DOE or the U.S. Government. The U.S. Government retains and the publisher, by accepting the article for publication, acknowledges that the U.S. Government retains a nonexclusive, paid-up, irrevocable, worldwide license to publish or reproduce the published form of this work or allow others to do so, for U.S. Government purposes.